\documentclass[reprint,nofootinbib,groupedaddress,amsmath,amssymb,aps,onecolumn]{revtex4}

\usepackage{graphicx}
\usepackage{amssymb}
\usepackage{amsmath}
\usepackage{mathrsfs}
\usepackage{textcomp}
\usepackage{epsfig}
\usepackage{slashed}
\usepackage{comment}
\usepackage{tensor}
\usepackage{xcolor}

\renewcommand{\d}{\mathrm{d}}
\newcommand{\dd}{\partial}

\begin{document}

\title{Noncommutative Reissner--Nordstr\"om black hole from noncommutative charged scalar field}

\author{Marija Dimitrijevi\'c \'Ciri\'c}
\email{dmarija@ipb.ac.rs}
\affiliation{Faculty of Physics, University of Belgrade, Studentski trg 12, 11000 Beograd, Serbia}

\author{Tajron Juri\'c}
\email{tjuric@irb.hr}
\affiliation{Rudjer Bo\v skovi\'c Institute, Bijeni\v cka c.54, HR-10002 Zagreb, Croatia}

\author{Nikola Konjik}
\email{konjik@ipb.ac.rs}
\affiliation{Faculty of Physics, University of Belgrade, Studentski trg 12, 11000 Beograd, Serbia}

\author{Andjelo Samsarov}
\email{asamsarov@irb.hr}
\affiliation{Rudjer Bo\v skovi\'c Institute, Bijeni\v cka c.54, HR-10002 Zagreb, Croatia}

\author{Ivica Smoli\'c}
\email{ismolic@phy.hr}
\affiliation{Theoretical Physics Department, Faculty of Science, University of Zagreb, p.p. 331, HR-10002 Zagreb, Croatia}

\date{\today}

\begin{abstract}
Within the framework of noncommutative (NC) deformation of gauge field theory by the angular twist, we
first  rederive the NC scalar and gauge field model from our previous papers and then  generalize
it to the second order in the Seiberg-Witten (SW) map. It turns out that SW expansion is finite and that it ceases at the second order in the deformation parameter, ultimately giving rise to the equation of motion for the scalar field in Reissner--Nordstr\"om (RN) metric that is nonperturbative and exact at the same order.
As a further step, we show  that the effective metric put forth and constructed in our previous work satisfies the equations of Einstein-Maxwell gravity, but only within  the first order of deformation and when the gauge field is fixed   by the Coulomb potential of the charged black hole. Thus obtained NC deformation of the Reissner--Nordstr\"om (RN) metric appears to have an additional off-diagonal element which scales linearly with a deformation parameter. We analyze  various properties of this metric.

\end{abstract}

\maketitle

\section{Introduction}

So far  general relativity (GR) has  shown to be  a highly successful theory of gravity, which was manifested in its remarkable ability to explain a  series of observations \cite{1,2,3,4} ranging from the early days examinations of the perihelion precession of Mercury, the bending of
light and the gravitational redshift of radiation from distant stars to modern day experimental achievements in detecting the gravitational waves
and imaging of black holes. What was in a not so recent past only a  mere theoretical conception, 
 by the appearance of advanced ground-based and space-based missions \cite{5,6,7,8,9} like the LIGO and the Event Horizon Telescope, soon became a factual physical reality.
While the LIGO experiment  set the ground for the first ever detection of gravitational waves by the colliding
black holes and neutron stars,  the Event Horizon Telescope provided the image of the black hole M87* (actually the image of the gas orbiting around the black hole at the center of the supergiant elliptical galaxy Messier 87), thus  further adding to a GR's enviable predictive power \cite{10,11,12,13,14,15,16}. 

However, in order that a premise of general relativity as  the correct theory of gravity be sustained, it was necessary to introduce into a consideration  few exotic ingredients such as the  
dark matter and the dark energy \cite{20,21,22}, to explain the
galactic rotation curves and the accelerated expansion of the universe. 
Besides,  the conceptual problems  with the black hole and the Big Bang singularities \cite{17,18} point to the fact that  the ultraviolet character of gravity still lacks a complete understanding. With all these issues, any attempt  to modify  general relativity or to consider  alternative gravity models appears to come as a  quite natural endeavor.

In this paper we use the methods of noncommutative geometry and 
noncommutative gravity \cite{nc,nc1,nc2,nc3,nc4,nc5,nc6,Aschieri:2005yw,nc8}  first to recapitulate
a construction of the NC scalar and gauge field model from the reference \cite{rnqnm} and then to generalize
it to the second order in the Seiberg-Witten map \cite{swmap}. It will appear that the second order in the expansion is at the same time
an ultimate order and consequently the model obtained is nonperturbative and exact.
Using  certain duality symmetries that are present at the first order in SW expansion, we  recap 
a construction from \cite{Dimitrijevic-Ciric:2022vvl,rnqnm} that gives rise to a particular noncommutative deformation of the RN metric.
While this construction turns out to be possible at the first order in SW, it  fails at the second order.
This is due to the  fact that duality symmetry breaks at the second order in SW expansion.

As a further step, we show  that the metric put forth and constructed in  \cite{Dimitrijevic-Ciric:2022vvl} appears to be a deformation of the Reissner--Nordstr\"om (RN) black hole that acquires an additional off-diagonal element, linear in the deformation parameter,  and satisfying the Einstein-Maxwell equations at the first order of deformation.
The construction in reference \cite{Dimitrijevic-Ciric:2022vvl} was carried out by utilizing the methods of noncommutative (NC) gauge-field theory \cite{nc,nc1,nc2,nc3,nc4,nc5,nc6,Aschieri:2005yw,nc8}  coupled to NC spinor field and to a  classical geometry of the RN type.
 The methods of NC gauge theory and gravity offer a yet another convenient way to modify general relativity in order to capture effects that are expected to appear  close to the  Planck scale. The ultimate hope is that the NC modifications  of gravity will unravel something of its  quantum character.
 In the rest of the paper we then go on to explore the physical properties of this NC deformed metric and try to understand its origin and meaning.



It is noteworthy that the construction
 considered in \cite{Dimitrijevic-Ciric:2022vvl} was not the only attempt
in the literature to deform the  RN metric (within the framework of noncommutative physics). Indeed,
in recent years there have been several investigations concerning the noncommutative versions of  Reissner-Nordstr\"om (RN) black hole \cite{Nozari:2007ya, Alavi:2009tn, Romero-Ayala:2015fba, Spallucci:2009zz, Mukherjee:2007fa, Bufalo:2014bwa}.
Most of the research in the literature however was dealing with the  so-called Moyal-type noncommutativity $[\hat{x}^{\mu},\hat{x}^{\nu}]=\theta^{\mu\nu}$. For example, in \cite{Nozari:2007ya, Alavi:2009tn, Spallucci:2009zz} the authors have used this type of noncommutativity and implemented it using the smeared $\delta$-functions for the mass and charge distribution. The main feature of such systems are the change in Hawking temperature and entropy.  An alternative approach has been presented in \cite{Romero-Ayala:2015fba} where the Moyal noncommutativity was introduced using deformed embedding of RN into deformed Riemannian geometry. Using the  framework of NC gauge theory of gravity authors of \cite{Mukherjee:2007fa} were able to construct corrections to RN solution and showed that this could lead to a removal of singularities. \\

The structure of the paper is as follows: In the following section we review our model of the NC charged scalar field in a curved background coupled to the NC $U(1)$ gauge field. In Section III we extend our results to the second order expansion in the deformation parameter and show that the equation of motion for the NC scalar field does not contain higher order terms. This defines the exact (in the NC parameter expansion) model of a NC charged scalar field coupled to the curved (spherically symmetric) background. In Section IV and V we discuss the properties of the NC charged black hole, obtained from the effective/dual metric in the equation of motion for the NC scalar filed. Section VI contains further discussion and some conclusions. In particular, we comment on the possibility to construct the effective metric up to the second order in the deformation parameter by introducing the nonmetricity tensor.

\section{NC scalar field in the Reissner-Nordstr\"om background}

Consider a system consisting of a charged scalar and $U(1)$ gauge field, as well as of classical gravitational field.
We want to deform this system in order to ultimately generate one type of  deformation of the 
classical solution to the Einstein's gravity. In particular, it is a noncommutative deformation of the RN metric that we
focus our attention to.
The required steps may be carried out by
following \cite{Dimitrijevic-Ciric:2022vvl}, where the noncommutativity was introduced at the level of scalar field that is probing the underlying RN background. After  a careful derivation of the corresponding equation of motion, which we briefly repeat here, one comes to a conclusion that this system of RN black hole coupled to a NC scalar and gauge fields is equivalent/dual to a system of a commutative scalar field and a new effective background, which up to the first order in the deformation parameter absorbs all NC effects. We refer to this effective background as the noncommutative Reissner--Nordstr\"om (NCRN) black hole.

Let us start with writing the action functional for the NC $U(1)_\star$ gauge theory of a massless charged scalar field $\hat{\phi}$ in an arbitrary background (that has Killing vectors $\partial_t$ and $\partial_{\varphi}$) \cite{rnqnm}
\begin{equation}\begin{split}\label{action}
\mathcal{S}[\hat{\phi},\hat{A}]&=\int (d\hat{\phi}-i\hat{A}\star\hat{\phi})^\dagger\wedge_{\star}*(d\hat{\phi}-i\hat{A}\star\hat{\phi})-\frac{1}{4}\int\hat{F}\wedge_{\star}*\hat{F}\\
&=\int d^4 x\sqrt{\left|g\right|}\star g^{\mu\nu}\star D_{\mu}\hat{\phi}^\dagger \star D_{\nu}\hat{\phi}-\frac{1}{4}\int d^4 x\sqrt{\left|g\right|}\star g^{\alpha\beta}\star g^{\mu\nu}\star\hat{F}_{\alpha\mu}\star\hat{F}_{\beta\nu}
\end{split}\end{equation}
where $D_{\mu}$ is the  covariant derivative defined by 
\begin{equation}
D_{\mu}\hat{\phi}=\partial_{\mu}\hat{\phi}-iq\hat{A}\star\hat{\phi}
\end{equation}
and $\hat{F}=\hat{F}_{\mu\nu}\star dx^{\mu}\wedge_{\star}dx^{\nu}$ is the field-strength defined by
\begin{equation}
\hat{F}_{\mu\nu}=\partial_{\mu}\hat{A}_{\nu}-\partial_{\nu}\hat{A}_{\mu}-i[\hat{A}_{\mu}\overset{\star}{,}\hat{A}_{\nu}].
\end{equation}
Tha action (\ref{action}) is written in spherical coordinates $x^\mu = (t,r,\theta,\varphi)$ and the Hodge dual is denoted by $*$.

The $\star$-product is given by an Abelian twist 
\begin{equation}\label{twist}
\mathcal{F}=e^{-\frac{ia}{2}(\partial_t\otimes\partial_\varphi-\partial_\varphi\otimes\partial_t)}=e^{-\frac{i}{2}\theta^{\alpha\beta}\partial_{\alpha}\otimes\partial_\beta}
\end{equation}
via\footnote{$m$ is the multiplication map $m(a\otimes b)=ab.$}
\begin{equation}\begin{split}\label{star}
f\star g&=m\left(\mathcal{F}^{-1}\triangleright f\otimes g\right)\\
&=fg+\frac{ia}{2}\left(\frac{\partial f}{\partial t}\frac{\partial g}{\partial \varphi}-\frac{\partial f}{\partial \varphi}\frac{\partial g}{\partial t}\right)+\mathcal{O}(a^2),
\end{split}\end{equation}
where $f, g \in \mathcal{C}^{\infty}$ and
$\theta^{\alpha\beta}$ are 
  components of the NC deformation, with only $\theta^{t\varphi}$ and $\theta^{\varphi t}$ being different from zero, $\theta^{t\varphi}=-\theta^{\varphi t}=a.$  Note that this twist leads to the only nonvanishing commutator $[ t \overset{\star} {,}  e^{i\varphi} ] = -ae^{i\varphi}.  $
The twist (\ref{twist}) may be seen as a special case of the general class of  twists related to the Lie-algebraic deformation of Minkowski space \cite{Meljanac:2017oek}.
Note that since the twist $\mathcal{F}$ acts trivially on the metric, the $\star$-product in $\sqrt{\left|g\right|}\star g^{\alpha\beta}\star g^{\mu\nu}$ can be omitted. Now it is straightforward to check that the action \eqref{action} is invariant under the infinitesimal $U(1)_{\star}$ gauge transformations defined by
\begin{equation}
\delta_{\star}\hat{\phi}=i\hat{\Lambda}\star\hat{\phi}, \quad \delta_{\star}\hat{A}_{\mu}=\partial_{\mu}\hat{\Lambda}+i[\hat{\Lambda}\overset{\star}{,}\hat{A}_{\mu}], \quad \delta_{\star}\hat{F}_{\mu\nu}=i[\hat{\Lambda}\overset{\star}{,}\hat{F}_{\mu\nu}]
\end{equation}
where $\hat{\Lambda}$ is the NC gauge parameter.

Using the Seiberg-Witten(SW) map \cite{swmap} one can express the NC fields as functions of the corresponding commutative fields, which can then be expanded as a series in the deformation parameter $a$. Using the twist \eqref{twist}, one obtains the following recursion relations \cite{paololeonardo}
\begin{equation}\begin{split}\label{sw}
\hat{\phi}^{(n+1)} &= -\frac{1}{4(n+1)}\theta^{\rho\sigma}\Big( \hat{A}_{\rho}\star(\partial_{\sigma}\hat{\phi} + D_{\sigma}\hat{\phi})\Big)^{(n)},  \\
\hat{A}^{(n+1)}_{\mu} &= -\frac{1}{4(n+1)}\theta^{\rho\sigma} \Big( \{ \hat{A}_{\rho} \overset{\star}{,} (\partial_\sigma \hat{A}_\mu + \hat{F}_{\sigma\mu})\} \Big)^{(n)}, \\
\hat{F}^{(n+1)}_{\mu\nu} &= -\frac{1}{4(n+1)}\theta^{\rho\sigma}\Big( \{
{\hat A}_\rho \overset{\star}{,}
\partial_\sigma {\hat F}_{\mu\nu} + D_\sigma {\hat F}_{\mu\nu} \} \Big)^{(n)}
+\frac{1}{2(n+1)}\theta^{\rho\sigma}\Big( \{ {\hat F}_{\mu\rho} \overset{\star}{,}
{\hat F}_{\nu\sigma} \}
\Big)^{(n)}.
\end{split}\end{equation}
Using the first order results of \eqref{sw} and the $\star$-product \eqref{star} we expand the action \eqref{action} up to the first order in the deformation parameter $a$ as follows
\begin{equation}\begin{split}\label{actiona}
\mathcal{S}&=\int d^4x\sqrt{\left|g\right|}\Big(D_{\mu}\phi^\dagger D^{\mu}\phi-\frac{1}{4}F_{\mu\nu}F^{\mu\nu}+\frac{1}{8}g^{\mu\rho}g^{\nu\sigma}\theta^{\alpha\beta}(F_{\alpha\beta}F_{\mu\nu}F_{\rho\sigma}-4F_{\mu\alpha}F_{\nu\beta}F_{\rho\sigma})\\
& +\frac{1}{2}\theta^{\alpha\beta}g^{\mu\nu}(-\frac{1}{2}F_{\alpha\beta}D_\mu \phi^\dagger D_{\nu}\phi +F_{\alpha\nu}D_\mu\phi^\dagger D_{\beta}\phi +F_{\alpha\mu}D_\beta\phi^\dagger D_{\nu}\phi )\Big)+\mathcal{O}(a^2),
\end{split}\end{equation}
where $D_{\mu}$ is the usual $U(1)$ covariant derivative $D_{\mu}\phi=\partial_{\mu}\phi-iqA_{\mu}\phi$. 
If we add the classical EH action to (\ref{actiona}), the resulting functional may be viewed as a deformation of Einstein-Maxwell gravity leading to an effective theory of gravity akin to
some effective models of gravity obtained in the low-energy limit of a string theory action containing the gravitational, gauge and dilaton or axion fields
\cite{Banerjee:2020ubc,Banerjee:2020qmi}.\\

By varying the action \eqref{actiona} with respect to $\phi^\dagger$ one obtains an equation of motion for $\phi$
\begin{eqnarray}\label{eom}
  && g^{\mu\nu}  \Big[D_{\mu}D_{\nu}\phi-\Gamma^{\lambda}_{\mu\nu}D_{\lambda}\phi  \nonumber  \\   
&-& \frac{1}{4}\theta^{\alpha\beta}\big( D_\mu(F_{\alpha\beta}D_\nu\phi)-\Gamma^{\lambda}_{\mu\nu}F_{\alpha\beta}D_\lambda\phi-2D_\mu(F_{\alpha\nu}D_\beta\phi)+2\Gamma^{\lambda}_{\mu\nu}F_{\alpha\lambda}D_\beta\phi-2D_\beta(F_{\alpha\mu}D_\nu\phi) \big)\Big]=0.
\end{eqnarray}
Varying the action with respect to $A_\lambda$ one can obtain the NC Maxwell's equations \cite{rnqnm}.

\smallskip

Gravitational background is defined by the Reissner--Nordstr\"om spacetime, with metric
\begin{equation}\label{rn}
g_{\mu\nu}=\begin{pmatrix}
-f(r)&0&0&0\\
0&\frac{1}{f(r)}&0&0\\
0&0&r^2&0\\
0&0&0&r^2 \sin^2\theta\\
\end{pmatrix}, \qquad f = 1 - \frac{2M}{r} + \frac{Q^2}{r^2} \, ,
\end{equation}
where $M$ is the mass and $Q$ the charge of the RN black hole, and the $U(1)$ gauge field
\begin{equation}\label{A}
A_{\mu} = (-\frac{Q}{r}, \, \vec{0} \, ) .
\end{equation}
Corresponding field-strength $F_{\mu \nu}$ has the only nonvanishing components 
\begin{equation}\label{F}
F_{tr} = - F_{rt} = -\frac{Q}{r^2}.
\end{equation}
Furthermore, since the only non-vanishing components of the NC deformation $\theta^{\alpha\beta}$ are $\theta^{t\varphi}=-\theta^{\varphi t}=a,$ by inserting \eqref{rn}, \eqref{A} and \eqref{F} into \eqref{eom} we finally obtain
\begin{equation}\label{eomc}
\left(\frac{1}{f}\partial^{2}_{t}-\Delta+(1-f)\partial^{2}_{r}+\frac{2M}{r^2}\partial_r +\frac{2iqQ}{rf}\partial_t-\frac{q^2 Q^2}{r^2f}\right)\phi+\frac{aqQ}{r^3}\left(\left(\frac{M}{r}-\frac{Q^2}{r^2}\right)\partial_\varphi+rf\partial_r\partial_\varphi\right)\phi=0.
\end{equation}
Equation \eqref{eomc} is the equation of motion of a NC scalar field in the background of the RN black hole. This equation, its quasinormal mode solutions and the Bekenstein-Hawking entropy  were extensively studied in \cite{rnqnm, DimitrijevicCiric:2019hqq, Gupta:2022oel}. Note that in the limit $a\to 0$ one obtains the usual equation of motion of a commutative scalar field in the RN background.

\section{Exact equation in the second order SW map}

In this section we extend the previous analysis to the second order in the SW expansion.
Remarkably, the SW expansion will terminate at this order and consequently the resulting equations
of motion will be exact. In order to find the second order NC corrections, we will use recurrent relations for SW map \eqref{sw} and follow steps similar to those in Section 3 and Section 4 in \cite{paololeonardo}. Similarly to \eqref{sw}, the recursion relations for the action \eqref{action} allow us to express corrections in order $n+1$ from the corrections in order $n$ by substituting all pointwise products with the $\star$-products and commutative fields with the corresponding NC fields. By closer inspection and taking into account that only nonvanishing components of the $\theta^{\mu\nu}$ and $F_{\mu\nu}$ are $\theta^{t\varphi}$ and $F_{tr}$, we see that the terms from the first order expansion which give nonzero corrections in the second order are the following\footnote{The superscript in $(...)^{(i)}$ denotes that only the $i$-th order in the deformation parameter $a$ is retained.}:
\begin{equation}\label{sw1}
  S^{(2)}= \sqrt{|g|} \frac{\theta^{\alpha\beta}g^{\mu\nu}}{2}(D_\mu\hat{\phi}^\dagger\star \hat{F}_{\alpha\nu}\star D_\beta\hat{\phi}+D_\beta\hat{\phi}^\dagger\star \hat{F}_{\alpha\mu}\star D_\nu\hat{\phi})^{(1)}.
\end{equation}
Inserting the SW map solutions \eqref{sw} and expanding the $\star$-products, with a help of the useful method for obtaining manifestly covariant results, given in Appendix B in  \cite{vojamarija}, \eqref{sw1} becomes 
\begin{eqnarray}
S^{(2)}=&&\sqrt{|g|}\frac{1}{4}\theta^{\alpha\beta}\theta^{\gamma\delta}g^{\mu\nu}(-2A_\gamma\partial_\delta(D_\mu\phi^\dagger F_{\alpha\nu}D_\beta\phi)+iD_\gamma (D_\mu\phi^\dagger D_\beta\phi)D_\delta F_{\alpha\nu}\nonumber\\
&&+iF_{\alpha\nu}(D_\gamma D_\mu\phi^\dagger)(D_\delta D_\beta\phi)+D_\mu\phi^\dagger F_{\alpha\nu}F_{\gamma\beta}  D_\delta\phi\nonumber\\
&&+ D_\delta\phi^\dagger F_{\alpha\nu}F_{\gamma\mu}D_\beta\phi + 2D_\mu\phi^\dagger F_{\gamma\alpha}F_{\delta\nu}D_\beta\phi)\nonumber\\
&&+\sqrt{|g|}\frac{1}{4}\theta^{\alpha\beta}\theta^{\gamma\delta}g^{\mu\nu}(-2A_\gamma\partial_\delta(D_\beta\phi^\dagger F_{\alpha\mu}D_\nu\phi)+iD_\gamma (D_\beta\phi^\dagger D_\nu\phi)D_\delta F_{\alpha\mu}\nonumber\\
&&+iF_{\alpha\mu}(D_\gamma D_\beta\phi^\dagger)(D_\delta D_\nu\phi)+ D_\beta\phi^\dagger F_{\alpha\mu}F_{\gamma\nu}  D_\delta\phi \nonumber\\
&&+D_\delta\phi^\dagger F_{\alpha\mu}F_{\gamma\nu}D_\nu\phi+2D_\beta\phi^\dagger F_{\gamma\alpha}F_{\delta\mu}D_\nu\phi).
\end{eqnarray}
After subsequent partial integrations and the use of the identity $i[D_\alpha,D_\beta]\phi=F_{\alpha\beta}\phi,$ as well as the fact that  derivatives which are contracted with the NC deformation parameter matrix $\theta^{\alpha\beta}$ do not act on the field strength tensor $F_{\mu\nu}$, we obtain
\begin{eqnarray}
S^{(2)}=&&\sqrt{|g|}\frac{1}{4}\theta^{\alpha\beta}\theta^{\gamma\delta}g^{\mu\nu}(-F_{\gamma\delta}(D_\mu\phi^\dagger F_{\alpha\nu}D_\beta\phi)+iD_\gamma (D_\mu\phi^\dagger D_\beta\phi)D_\delta F_{\alpha\nu}\nonumber\\
&&-F_{\alpha\nu}(D_\mu\phi^\dagger)F_{\gamma\delta}(D_\beta\phi)+D_\mu\phi^\dagger F_{\alpha\nu}F_{\gamma\beta}  D_\delta\phi\nonumber\\
&&+ D_\delta\phi^\dagger F_{\alpha\nu}F_{\gamma\mu}D_\beta\phi +2D_\mu\phi^\dagger F_{\gamma\alpha}F_{\delta\nu}D_\beta\phi)\nonumber\\
&&+\frac{1}{4}\sqrt{|g|}\theta^{\alpha\beta}\theta^{\gamma\delta}g^{\mu\nu}(-F_{\gamma\delta}(D_\beta\phi^\dagger F_{\alpha\mu}D_\nu\phi)+iD_\gamma (D_\beta\phi^\dagger D_\nu\phi)D_\delta F_{\alpha\mu}\nonumber\\
&&+F_{\alpha\mu}( D_\beta\phi^\dagger)F_{\gamma\delta}( D_\nu\phi)+ D_\beta\phi^\dagger F_{\alpha\mu}F_{\gamma\nu}  D_\delta\phi\nonumber\\
&&+D_\delta\phi^\dagger F_{\alpha\mu}F_{\gamma\beta}D_\nu\phi+2D_\beta\phi^\dagger F_{\gamma\alpha}F_{\delta\mu}D_\nu\phi).
\end{eqnarray}
Since $F_{t\varphi}=0$, some of the above terms vanish, while the others add to one term given by
\begin{equation}
  S^{(2)}= \sqrt{|g|} \frac{1}{4}\theta^{\alpha\beta}\theta^{\gamma\delta}g^{\mu\nu}(D_\beta\phi^\dagger F_{\alpha\mu}F_{\gamma\nu}  D_\delta\phi+D_\beta\phi^\dagger F_{\alpha\mu}F_{\gamma\nu}  D_\delta\phi)
   = \sqrt{|g|}\frac{1}{2}\theta^{\alpha\beta}\theta^{\gamma\delta}g^{\mu\nu}(D_\beta\phi^\dagger F_{\alpha\mu}F_{\gamma\nu}  D_\delta\phi).
\end{equation}
Variation of these terms with respect to  $\phi^\dagger$ gives rise to additional terms in the equation of motion.
It turns out that there appears only one new term, which is of the form
\begin{equation}
\frac{1}{2}\theta^{\alpha\beta}\theta^{\gamma\delta}g^{\mu\nu}F_{\alpha\mu}F_{\gamma\nu}D_\beta D_\delta \phi .\nonumber
\end{equation}
More explicitly, we obtain
\begin{equation}
\frac{1}{2}\theta^{t\varphi}\theta^{t\varphi}g^{rr}F_{tr}F_{tr}\partial^2_\varphi\phi = \frac{1}{2}a^2 (-f)\frac{q^2Q^2}{r^4}\partial^2_\varphi\phi = -\frac{a^2q^2Q^2}{2r^4}f\partial^2_\varphi\phi . \nonumber
\end{equation}
Finally, the resulting equation of motion is
\begin{eqnarray}
&&
\Big( \frac{1}{f}\partial^2_t -\Delta + (1-f)\partial_r^2 
+\frac{2MG}{r^2}\partial_r + 2iqQ\frac{1}{rf}\partial_t -\frac{q^2Q^2}{r^2f}\Big)\phi
\nonumber\\
&& +\frac{aqQ}{r^3}
\Big( (\frac{MG}{r}-\frac{GQ^2}{r^2})\partial_\varphi
+ rf\partial_r\partial_\varphi \Big) \phi-\frac{a^2q^2Q^2}{2r^4}f\partial^2_\varphi\phi =0 . \label{2ndorderequation}
\end{eqnarray}

As already noted, the equation of motion (\ref{2ndorderequation}) is not just a perturbative result valid up to the second order in  deformation. It is an exact result and  may be attributed to the SW map terminating
at that same order. As an advantageous  outcome, one finds that all analysis that is ever going to follow from this equation
will  require no perturbative protocols  anymore. All results following from (\ref{2ndorderequation}) are going to be exact automatically. There is one more way to justify why the equation (\ref{2ndorderequation}) is exact and no higher order corrections appear. Namely, the SW map is linear in matter fields, while the action \eqref{action} is quadratic in the matter field $\phi$. The only nonzero components of the deformation parameter $\theta$ are $\theta^{t\varphi}$, so each new order of expansion will contribute one additional set of $\varphi$ and $t$ indices. Note that the index $\varphi$ can only appear contracted to $D_\varphi\phi$, since all $F_{\mu\varphi}=0$. Since the action \eqref{action} is quadratic in the field $\phi$ and we can always partially integrate multiple covariant derivatives on $\phi$ to obtain $F_{\rho\sigma}$, we conclude that the maximal number of $D_\varphi$ in the expanded action is two and therefore the expansion of the action has to terminate at the second order.

\section{Noncommutative Reissner-Nordstr\"om black hole}

In this section we focus on the first order in the SW expansion, that is the equation of motion (\ref{eomc}), and identify a duality symmetry that exists at that order.
This symmetry will alow us to absorb the noncommutative contributions into a single d'Alembertian operator and ultimately to identify the effective metric related to this problem,
which will turn out to be a deformation of the Reissner--Nordstr\"om metric. We will later discuss  possible extensions of the duality symmetry to higher orders.

The equation of motion for the NC scalar field minimally coupled to the RN background can be written in the following form \cite{Dimitrijevic-Ciric:2022vvl}
\begin{equation}\label{E}
\frac{1}{\sqrt{\left|g\right|}} \, D_{\mu}(\sqrt{\left|g\right|}g^{\mu\nu}D_{\nu}\phi)+\Box_a \phi=0,
\end{equation}
where $\Box_a$ is the part of \eqref{eomc} which contains only NC contributions.  Now, we try to rearrange \eqref{E} so that  the NC operator $\Box_a$ is absorbed  into some effective metric $g^{\prime}_{\mu\nu}$. Namely, we write 
\begin{equation}\label{dd}
\frac{1}{\sqrt{\left|g^\prime\right|}} \, D_{\mu}(\sqrt{\left|g^\prime\right|}g^{\prime\mu\nu}D_{\nu}\phi) = \frac{1}{\sqrt{\left|g\right|}} \, D_{\mu}(\sqrt{\left|g\right|}g^{\mu\nu}D_{\nu}\phi) + \Box_a \phi.
\end{equation}
We can write an ansatz for $g^{\prime}_{\mu\nu}$ and after carefully comparing the left and right hand sides of \eqref{dd}, one can extract the components of the effective metric $g^{\prime}_{\mu\nu}$ to obtain
\begin{equation}  \label{1stordereffmetric}
g^{\prime}_{\mu\nu}=\begin{pmatrix}
-f&0&0&0\\
0&\frac{1}{f}&0&\frac{aqQ}{2}\sin^2\theta\\
0&0&r^2&0\\
0&\frac{aqQ}{2}\sin^2\theta&0&r^2 \sin^2\theta\\
\end{pmatrix}+\mathcal{O}(a^2).
\end{equation}
Since we have an effective metric $g^{\prime}_{\mu\nu},$ we can notice an equivalence between the equation of motion of a NC scalar field in the RN background \eqref{eom} and the equation of motion guiding a commutative scalar field on some effective background endowed with the effective metric $g^{\prime}_{\mu\nu}$. The similar property has already been observed for the NC scalar field
on the BTZ background \cite{ncbtz, ncbtz1, ncbtz2, ncbtz3}  in the context of $\kappa$-deformation \footnote{In particular, in \cite{ncbtz, ncbtz1, ncbtz2, ncbtz3} it was shown that noncommutativity may give rise to the black hole spin and that it essentially mimicks its advent. It is interesting to note that a similar type of  feature, where the noncommutativity is assigned a role of a mimicker of some specific physical property, is quite usual in the literature, see for example the reference \cite{Nozari:2006rt}.}.  As this effective metric 
$g^\prime_{\mu\nu}$  appears to absorb all  NC effects,  we will name this new effective space as NCRN and in what follows we will investigate its physical properties. This effective metric provides a dual picture to the same physical system, comprising of the NC scalar field with the charge $q$ and the background metric generated by the black hole with mass $M$ and charge $Q.$ Note that from now on we will be dealing with only one metric, that pertaining to NCRN and for simplicity we switch the notation accordingly, i.e. $g^\prime \to g.$ 

\smallskip

Thus, the metric of NCRN is given by
\begin{equation}\label{ncrn}
g_{\mu\nu}=\begin{pmatrix}
-f(r) & 0 & 0 & 0 \\
0 & \frac{1}{f(r)} & 0 & A\sin^2\theta \\
0 & 0 & r^2 & 0 \\
0 & A\sin^2\theta & 0 & r^2 \sin^2\theta \\
\end{pmatrix},
\end{equation}
written with the abbreviations
\begin{equation}
f(r) = 1 - \frac{2M}{r} + \frac{Q^2}{r^2} \qquad \textrm{and} \qquad A = \frac{aqQ}{2} \, .
\end{equation}
As $A\to 0$ we recover the commutative limit. 
Interestingly, when the same procedure is carried out for the spin $1/2$ field up to  first order in deformation, the same effective metric (\ref{ncrn}) arises \cite{Dimitrijevic-Ciric:2022vvl}. Situation with the extension of this analysis to the vector field is however little bit different.
Namely, for the electromagnetic spin 1 field there are no corrections
 to the equation of motion in the first order, while in higher orders
 in $\Theta$ due to SW-map \eqref{sw} the NC Maxwell equation becomes
 nonlinear in $A_\mu$, rendering any possibility of constructing a
 dual picture with an effective metric impossible.
On the other hand, in order to extend this construction to the second order in the deformation parameter, we need to allow a more general connection. We comment on this in the concluding section.

From now on, we will drop scalar field from any subsequent discussion and the only subject of our interest will be a system consisting of the gauge field and the gravitational field (metric tensor).

\smallskip


\smallskip

Now, the main question is what  geometry and physics do lie behind the NCRN metric \eqref{ncrn}. Let us evaluate the Einstein tensor
\begin{equation}\label{einstein}
G_{\mu\nu} = R_{\mu\nu} - \frac{1}{2} \, R g_{\mu\nu} = \begin{pmatrix}
\frac{Q^2 f}{r^4} & 0 & 0 & 0 \\
0 & -\frac{Q^2}{r^4 f} & 0 & A\frac{Q^2\sin^2 \theta}{r^4} \\
0 & 0 & \frac{Q^2}{r^2} & 0 \\
0 & A\frac{Q^2\sin^2 \theta}{r^4} & 0 & \frac{Q^2\sin^2\theta}{r^2}
\end{pmatrix} + \mathcal{O}(A^2) \, .
\end{equation}
As can be seen, the Einstein tensor is non-zero so that the NCRN metric, as expected, is not a vacuum solution to the Einstein equation.

 Thus, the NC effect may be encripted  within some effective matter source, appearing on the right hand side of the Einstein field equation.  The interesting feature is that for the metric (\ref{ncrn})  this effective matter source may be fixed by 
 the Maxwell's energy-momentum tensor,
\begin{equation}\label{maxtensor}
T^M_{\mu \nu} = \frac{1}{4\pi} \left( F_{\mu \lambda} \tensor{F}{_\nu^\lambda} - \frac{1}{4} g_{\mu \nu} F_{\lambda \sigma} F^{\lambda \sigma} \right) .
\end{equation}
Indeed, it may be shown that  up to first order in the deformation $A$ the metric (\ref{ncrn}) satisfies the Einstein-Maxwell field equation
\begin{equation}\label{magnmon}
G_{\mu \nu} = 8\pi T^M_{\mu \nu}. 
\end{equation}
We first note that the zeroth order in $A$ in (\ref{einstein}), i.e. the Einstein tensor for the RN metric, is proportional to the  Maxwell energy-momentum tensor (\ref{maxtensor}),
where the only non-vanishing component of the electromagnetic tensor $F_{\mu\nu}$ is $F_{rt}=-F_{tr}= Q/r^2.$
 In order to see what happens in higher orders, in particular the first order in $A,$
 we absorb  the NC corrections appearing in (\ref{einstein}) into the energy-momentum tensor $T^M_{ab},$ and  
simultaneously allow the modifications in the electromagnetic tensor $F_{ab}$. In doing so we propose the following ansatz
\begin{equation}\label{ansatz}
F_{\mu\nu}=\begin{pmatrix}
0&-\frac{Q}{r^2}-AF_0&AF_1&AF_2\\
\frac{Q}{r^2}+AF_0&0&AF_3&AF_4\\
-AF_1&-AF_3&0&AF_5\\
-AF_2&-AF_4&-AF_5&0\\
\end{pmatrix}
\end{equation}
where $F_i=F_i(t,r,\theta,\varphi)$ are yet unknown functions.
Now, using  the Einstein tensor \eqref{einstein} calculated for the metric  (\ref{ncrn}) and the energy-momentum tensor \eqref{maxtensor}, evaluated  for the ansatz \eqref{ansatz}, we can calculate the difference tensor 
\begin{equation}\label{razlika}
G_{\mu\nu}-8\pi T^{M}_{\mu\nu}=\begin{pmatrix}
-\frac{AQf}{r^2}F_0&0&-\frac{2AQf}{r^2}F_3&-\frac{2AQf}{r^2}F_4\\
0&\frac{2AQ}{r^2f}F_0&-\frac{2AQ}{r^2f}F_1&-\frac{2AQ}{r^2f}F_2\\
-\frac{2AQf}{r^2}F_3&-\frac{2AQ}{r^2f}F_1&-2AQF_0&0\\
-\frac{2AQf}{r^2}F_4&-\frac{2AQ}{r^2f}F_2&0&-2AQF_0\sin^2\theta\\
\end{pmatrix}+\mathcal{O}(A^2).
\end{equation}
The only way that the above difference tensor  vanishes is  if 
\begin{equation}\label{uvjet}
F_0=F_1=F_2=F_3=F_4=0,
\end{equation}
leaving the function $F_5(t,r,\theta,\varphi)$ still arbitrary.
Thus, we see that  up to first order in $A$    the metric (\ref{ncrn}) satisfies the Einstein-Maxwell field equation
(\ref{magnmon}).   \\


An alternative perspective on this situation might be  that nonvanishing $G_{\mu \nu} $ in (\ref{einstein})
results from a modification of Einstein's gravitational field equation. 
In that case  we are interpreting all corrections as coming from the (NC) geometry part \cite{Harikumar:2006xf}-\cite{Borowiec:2016zrc}, i.e. as corrections to the left hand side
of the Einstein equation. In doing so, one would obviously fix the energy-momentum part and modify the Einstein tensor $G_{\mu \nu} \longrightarrow \hat{G}_{\mu \nu}$
according to
\begin{equation}  \nonumber
\hat{G}_{\mu \nu} = R_{\mu \nu} - \frac{1}{2} \, g_{\mu \nu} R + \mathcal{O}(A) .
\end{equation}

\section{Physical properties of NCRN} 

In the following we make a review of some general properties of the metric  \eqref{ncrn}.
Later on we shall see that much of these properties may be easily understood through the lenses
 of a  transition to another coordinates.


\subsection{Various aspects of NCRN} 

Primarily, it is easy to see that this metric is static since its stationary Killing vector field $k = \dd/\dd t$ satisfies $k \wedge \d k = 0$ and metric is written explicitly in block-diagonal form. Furthermore, by Vishveshwara-Carter's theorem we know that ergosurfaces, consisting of points where Killing vector field $k^a$ becomes null, coincides with the Killing horizon $H[k]$ generated by $k^a$.

\smallskip

On the other hand, the horizon can be quickly found by looking at the zeros of the metric function $f(r)$, which are formally identical as in the commutative Reissner-Nordstr\"om black hole. However, as the original coordinate system in which the metric is written is \emph{not} regular at the black hole horizon, we have to use some of the light-like coordinates, such as $v = t + r_*$ with the tortoise coordinate $r_*$ introduced via $\d r_* = \d r/f(r)$: spacetime metric in the coordinate system $\{v,r,\theta,\varphi\}$ takes the form
\begin{equation}
\d s^2 = -f(r)\d v^2 + 2 \, \d v \, \d r + 2A \sin^2\theta \, \d r \, \d\varphi + r^2 (\d\theta^2 + \sin^2\theta \, \d\varphi^2).
\end{equation}
Nevertheless, here we have $k = \dd/\dd v$ and again $k^2 = g_{vv} = -f(r)$. Let us denote the zeros of $f(r)$ with $r_+$ and $r_-$, so that 
$$r_\pm = M \pm \sqrt{M^2 - Q^2},$$
as in the case of RN black hole. 

\medskip

Another interesting point is the temperature of the NCRN black hole.
It appears that  the temperature of the Reissner-Nordstr\"om black hole remains unaltered by the noncommutative corrections. This may be seen from the following line of arguments, starting from the well known expression for the surface gravity,
\begin{equation}
\kappa^2 = -\lim_H \frac{(k^b \nabla_{\!b} k^a)(k^c \nabla_{\!c} k_a)}{k^a k_a}.
\end{equation}
The evaluation of the expression in the numerator gives
$$k^\alpha \nabla_{\!\alpha} k^\mu = k^\alpha \dd_\alpha k^\mu + k^\alpha \Gamma^\mu_{\alpha\beta} k^\beta = 0 + \Gamma^\mu_{tt} = \frac{1}{2}\,g^{\mu r} \dd_r f(r)$$
and consequently
\begin{equation}
(k^\alpha \nabla_{\!\alpha} k^\mu)(k^\beta \nabla_{\!\beta} k_\mu) = \frac{1}{4}\, g_{\mu\nu} g^{\mu r} g^{\nu r} (\dd_r f)^2 = \frac{1}{4}\, g^{rr} (\dd_r f)^2 \ , \quad k^2 = g_{tt} = -f.
\end{equation}
Since  the only component of the metric that we need is $g^{rr} $ and that it is given by
$$g^{rr} = \frac{1}{\frac{1}{f} - \frac{A^2 \sin^2\theta}{r^2}},$$
one finally gets
$$\kappa^2 = \lim_{r \to r_+} \frac{1}{4}\,\frac{(f')^2}{1 - \frac{A^2 \sin^2\theta}{r^2}\,f} = \lim_{r \to r_+} \frac{(f')^2}{4}.$$
Formally, as above, this has to be checked in a regular coordinate system, such as $\{v,r,\theta,\varphi\}$. Here we have
$$k^\alpha \nabla_{\!\alpha} k^\mu = \Gamma^\mu_{vv} = \frac{1}{2}\,g^{\mu r} \dd_r f(r)$$
and consequently,
$$(k^\alpha \nabla_{\!\alpha} k^\mu)(k^\beta \nabla_{\!\beta} k_\mu) = \frac{1}{4}\, g^{rr} (\dd_r f)^2 = \frac{1}{4}\,\frac{(f')^2}{\frac{1}{f} - \frac{A^2 \sin^2\theta}{r^2}}.$$
Again, the conclusion remains unaltered, $\kappa = f'(r_+)/2$. Expectedly, this result is in accordance with that obtained
in \cite{rnqnm} when calculating the emission rate of the scalar particles using the Parikh-Wilczek tunneling formalism. Moreover, the conclusion that the lowest nonvanishing NC correction to the horizon temperature is beyond the linear one seems to be in agreement with  other approaches in the literature \cite{Nozari:2008rc,Banerjee:2008gc,Nozari:2012bp}.

\subsection{The Newtonian limit}

The Newtonian limit is defined by the three following premises \cite{Carroll:1997ar}: 
\begin{enumerate}
\item particle is moving slowly with respect to the speed of light,
\item gravitational field is weak and can be considered as perturbation of a flat space,
\item gravitational field is static.
\end{enumerate}

The mathematical description of premise 1. is given by the requirement
\begin{equation}
\frac{dx^i}{d\tau}\ll \frac{dt}{d\tau},
\end{equation}
which simplifies the geodesic equation
\begin{equation} 
\frac{d^2 x^{\mu}}{d\tau^{2}}+\Gamma^{\mu}_{tt}\left(\frac{dt}{d\tau}\right)^2=0  \label{geo}.
\end{equation}
Moreover,  since the gravitational field is static, we have
$$\Gamma^{\mu}_{tt}=-\frac{1}{2}g^{\mu r}\partial_r g_{tt}. $$
In the subsequent analysis we will need the inverse of the metric \eqref{ncrn} which is given by 
\begin{equation}\label{inverz}
g^{\mu\nu}=\begin{pmatrix}
-\frac{1}{f}&0&0&0\\
0&f&0&-A\frac{f}{r^2}\\
0&0&\frac{1}{r^2}&0\\
0&-A\frac{f}{r^2}&0&\frac{1}{r^2 \sin^2\theta}\\
\end{pmatrix} +\mathcal{O}[A]^2.
\end{equation}
Now, if we examine the equation \eqref{geo} for $\mu=t,$ we get
\begin{equation}
\frac{d^2 t}{d\tau^2}=0 \quad \Longrightarrow \quad   \frac{dt}{d\tau}=\text{const.}
\end{equation}
which enables us to rewrite \eqref{geo} in terms of coordinate time $t$ only,
\begin{equation}
\frac{d^2 x^{\mu}}{d t^{2}} + \Gamma^{\mu}_{tt}=0 \, .
\end{equation}
At this point we use the premise 2., which tells us that the gravitational field is weak and that it can be treated as a perturbation of the flat metric. In fact, one is here dealing with two types of perturbations: gravitational and noncommutative. Therefore, the inverse of the metric can be written as
\begin{equation}
g^{\mu\nu}=\eta^{\mu\nu}-h^{\mu\nu}+Ak^{\mu\nu}+\mathcal{O}(A\cdot h, h^2, A^2),
\end{equation}
where only the lowest order in $h$ and $A$ is kept. Let us calculate the Christoffel symbol in this approximation,
\begin{equation}
\Gamma^{\mu}_{tt}=-\frac{1}{2}g^{\mu r}\partial_r g_{tt}=-\frac{1}{2}(\eta^{\mu r}-h^{\mu r}+Ak^{\mu r})\partial_r g_{tt}=-\frac{1}{2}(\eta^{\mu r}+Ak^{\mu r})\partial_r g_{tt} +\mathcal{O}[h^2, A^2].
\end{equation}
 In the last equality
 we used the fact that in the lowest order  $-g_{tt}=f(r)=1+\mathcal{O}[h]$, i.e. $\partial g\cong\mathcal{O}[h]$.
Thus the Christoffel symbols are
\begin{equation}
\Gamma^{t}_{tt}=\Gamma^{\theta}_{tt}=0, \quad \Gamma^{r}_{tt}=\frac{1}{2}\frac{\partial f}{\partial r}, \quad \Gamma^{\varphi}_{tt}=-\frac{A}{2r^2}\frac{\partial f}{\partial r}
\end{equation}
 so that \eqref{geo} in the Newtonian limit reduces to
\begin{equation}\label{eomn}
\ddot{r}=-\frac{1}{2}\frac{\partial f}{\partial r}, \quad \ddot{\varphi}=\frac{A}{2r^2}\frac{\partial f}{\partial r}, \quad \ddot{\theta}=0.
\end{equation}
While the noncommutativity doesn't affect the radial equation,  it affects the equation for the polar coordinate. Equations of motion \eqref{eomn} can be written in a unified way as
\begin{equation}\label{ncnewton}
\ddot{x}^i=-\tilde{\partial}_i V(r),
\end{equation}
where $V(r)=\frac{1}{2}f$ is the  generalized Newtonian potential\footnote{For all practical purposes it is really the Newtonian potential since $f\approx 1-\frac{2M}{r}$ for $\frac{2M}{r}\gg \frac{Q^2}{r^2}.$} and $\tilde{\partial}_i \equiv\partial_i + \tensor{\tilde{\Theta}}{_i^j} \partial_j$ is the generalized Laplacian with 
\begin{equation}
\tensor{\tilde{\Theta}}{_i^j} = \begin{pmatrix}
0&0&\frac{A}{2r^2}\\
0&0&0\\
-\frac{A}{2r^2}&0&0\\
\end{pmatrix}.
\end{equation}
Equation \eqref{ncnewton} represents the noncommutative version of the Newton equation.

\subsection{Geodesics in NCRN}

Let us investigate the geodesics for the classical, electrically neutral particle moving in the background of NCRN \eqref{ncrn}. For the sake of simplicity let us examine geodesics in $\theta = \pi/2$ plane. The $4$-velocity is $u^\mu = (\dot{t},\dot{r},0,\dot{\varphi})$, where dot denotes the derivative with respect to proper time (in case of the timelike geodesics), or with respect to some affine parameter (in case of null geodesics). Kinematics is encapsulated in the square of the 4-velocity,
\begin{equation}
-\kappa = u_\mu u^\mu = -f(r) \dot{t}^2 + \frac{\dot{r}^2}{f(r)} + r^2 \sin^2\theta \, \dot{\varphi}^2 + 2A \sin^2\theta \, \dot{r} \dot{\varphi} \, ,
\end{equation}
written with the parameter
\begin{equation}
\kappa = \left\{ \begin{array}{rl} 1 \ , & \textrm{timelike} \\ 0 \ , & \textrm{null} \end{array} \right.
\end{equation}
On the other hand, due to Killing vectors $k = \partial/\partial t$ and $m = \partial / \partial \varphi$, there are two conserved quantities: energy $e$ and angular momentum $\ell$,
$$e = -g_{\mu\nu} u^\mu k^\nu = f(r) \dot{t} \quad \Rightarrow \quad \dot{t} = \frac{e}{f(r)},$$
$$\ell = g_{\mu\nu} u^\mu m^\nu = A \sin^2\theta \, \dot{r} + r^2 \sin^2\theta \, \dot{\varphi} \quad \Rightarrow \quad \dot{\varphi} = \frac{\ell}{r^2 \sin^2\theta} - \frac{A}{r^2}\,\dot{r}.$$
Thus, taking into account that $\theta = \pi/2$ and noting that terms linear in $A$ cancel, we have 
$$-\kappa = \left( \frac{1}{f(r)} - \frac{A^2}{r^2} \right) \dot{r}^2 -\frac{e^2}{f(r)} + \frac{\ell^2}{r^2}.$$
Formally we can put this into standard form with an effective potential via auxiliary function $R(\tau)$, 
\begin{equation}
\frac{\dot{R}^2}{2}+V(r)=\frac{e^2}{2},
\end{equation}
where 
$$V(r)=\frac{f(r)}{r^2}(l^2+\kappa r^2), \quad \frac{\dot{R}^2}{f(r)} = \left( \frac{1}{f(r)} - \frac{A^2}{r^2} \right) \dot{r}^2 \quad \textrm{i.e.} \quad \d R = \d r \, \sqrt{1 - \frac{A^2}{r^2}\,f(r)} \ .$$
However, it is difficult to write explicitly this relation.

\smallskip

Interestingly, in this analysis  the circular trajectories ($\dot{r}=0$) are completely unaffected by noncommutativity. However,   a particle that would  be released from rest (i.e. with $\ell = 0$) at great distance from the black hole,  would nevertheless gain some non vanishing shift  in the angle  due to the  NC term $-A\dot{r}/r^2$. This would imply 
that the total time  of the free fall for the photon would display a difference when calculated and compared between commutative and noncommutative cases. Indeed,
if $l=0,$ we  see that the radial motion is unchanged ( up to $A^2$ it is the same situation as in commutative case) and unfolds according to
\begin{equation}
r(\tau)=R_0+e\tau,
\end{equation}
where $R_0$ is the initial radius $r(0)=R_0.$ The polar coordinate should then acquire the NC correction
\begin{equation}
\dot{\varphi}=-\frac{A}{r^2}\dot{r} \  \ \Longrightarrow \ \ \varphi(\tau)=\varphi_0-Ae\left(\frac{1}{R_0 e-e^2\tau}-\frac{1}{R_0 e}\right).
\end{equation}
However, one should be careful about proper interpretation of these results, in particular about observational claims. As far as it goes, in our analysis 
we were relying on a specific coordinate system, and in this particular case it  is not that intuitive as one would initially expect, this being due to a  presence of the $g_{r\varphi}$ component in the metric.  
For example, we could say that the experiment is performed by ``static observers'', that is observers with 4-velocities $u^a$ tangent to orbits of stationary Killing vector field $k^a = \dd_t^a$, more concretely $u^\mu = (1/\sqrt{-g_{tt}},0,0,0),$ in which case the conclusions drawn might be somewhat different. These issues will be addressed in more detail in the final section where we shall take on the task of finding a genuine physical interpretation of the NCRN metric \eqref{ncrn} and specifically of its only nonvanishing off-diagonal component $g_{r \varphi }$.

\section{Concluding remarks}

This work has provided a study of the noncommutative $U(1)$ gauge field-gravity model coupled to scalar field all up to the second order in the Seiberg-Witten map. If classical Einstein-Hilbert action
is added to this model,  the resulting setup may be viewed as a deformation of the system consisting of
the gauge field, gravitational field and dilaton field that one usually encounters in some models of quantum gravity after the low energy limit is being taken.
The approach that we use  provides yet another procedure to modify GR  in order to  make it more compatible  with physics that is expected to occur at the Planck scale.

Using duality symmetry that is present at the first order in SW, we have rederived the effective metric from the reference  \cite{Dimitrijevic-Ciric:2022vvl} (see the equation \eqref{ncrn}),
which turns out to be a noncommutative deformation of the Reissner--Nordstr\"om metric, with the only non-vanishing off-diagonal component sitting at the entry $(r, \varphi)$
and scaling linearly with deformation parameter $a.$ This metric has been shown to satisfy 
the equations of Einstein-Maxwell gravity when the gauge field is fixed 
 to be the Coulomb potential with origin in a black hole charge, albeit only within the first order of deformation.
On the contrary, as we demonstrate in the Appendix,  the construction of the effective metric fails at the second order in SW expansion, due to duality symmetry being broken there.
However, it is worthy to note that if 
we extend the definition of the connection and besides the ordinary Christoffels take it to
 involve also  the contorsion and nonmetricity,
then the construction of the effective metric can be pushed through all up to the second order and beyond. 
More precisely, it can be shown that the inverse of the effective metric that  in such extended framework is able to produce the exact (nonperturbative) equation  
(\ref{2ndorderequation})    by means of the general equation of motion      (\ref{dd}) appears to pick up an additional 
term in the component $g^{\varphi\varphi}$
\begin{equation} \label{gphiphi2a}
g^{\varphi\varphi(2)}=-\frac{a^2q^2Q^2}{2r^4}f.
\end{equation}
 Consequently, the effective metric itself in this more extended framework acquires corrections\footnote{Note that the inverse of the effective metric is given by the exact result, while its inverse, i.e.~the effective metric itself has been expanded up to second order in deformation.}
\begin{equation} \label{gphiphi2b}
g^{(2)}_{\varphi\varphi}=\frac{a^2q^2Q^2}{4}f \sin^4\theta, \quad  \quad
g^{(2)}_{rr}=-\frac{a^2q^2Q^2}{4r^2}\sin^2\theta.
\end{equation}
For details we refer the reader to the Appendix. Here we only make a notice that  such construction  is however nonunique.


In section IV we have touched upon an important question
dealing with  an  actual  interpretation and  better understanding  of the metric \eqref{ncrn},
which we here come back to. Specifically, we are interested in the
interpretation of the $g_{r\varphi}$ metric component, as well as the meaning of the coordinates in which the metric is expressed and  calculations  were carried out   previously (especially in the preceding section). A call for caution has already been released before, as there might be a possibility that the predictions obtained in a previous section might not be fully trusted, due to possible misinterpretation of the coordinates. Indeed, we should not jump into conclusion that the obtained results are completely reliable just because the coordinates we work with are denoted as standard spherical coordinates. In other words, just because some coordinate is denoted by ``$r$'' or ``$\varphi$'' does not mean automatically that they are ``usual spherical coordinates'' (e.g.~it might be that $r \in \left< -\infty,\infty \right>$ or even $\varphi \in \left< -\infty,\infty \right>$). For this purpose, here we  investigate this point  in some more detail.

\medskip



 Using new coordinates $(\tilde{t},\tilde{r},\tilde{\theta},\tilde{\varphi}) = (t,r,\theta,\varphi - Ar^{-1})$
the metric turns into
\begin{equation}\label{coordtr}
g_{\tilde{\mu} \tilde{\nu} } = \begin{pmatrix} -f(r) & & & \\ & h(r,\theta) & & \\ & & r^2 & \\ & & & r^2 \sin^2 \theta \end{pmatrix} \ , \quad h(r,\theta) = \frac{1}{f(r)} - \frac{(A\sin\theta)^2}{r^2}.
\end{equation}
Therefore, we see that upon making  a coordinate transformation, the NCRN metric  \eqref{ncrn} revamps in a new, more familiar format, which to a first order
of deformation appears to be no different than the RN metric. Moreover,
 in this coordinate system it is manifestly clear that the metric is asymptotically flat up to the first order in the NC deformation parameter. In addition, it can be easily checked that the same change of coordinates may be used to    transform away the deflection of the photon in its free fall toward the center of the black hole studied in the previous section, i.e. to erase the only seemingly nontrivial effect of the NCRN presented in this work.
This would consequently mean that the NC corrections present in the metric  \eqref{ncrn} are trivial and that they do not have any physical meaning.
In light of these findings it doesn't come as a surprise that the NCRN metric  \eqref{ncrn} appears to have  properties that we have so far encountered,
in particular that all nontrivial changes appear  at the order that is not lower than the second.

However, we want to stress  that  the above reasoning, as well as the  conclusions drawn from it  do not present the whole picture, but only a portion of it.
As such, this reasoning alone is insufficient to provide any reliable  or far-reaching conclusion and in many aspects is misleading.
It is indeed true that  
the metric tensor in the new coordinates at the first order in NC parameter $a$ 
 seems to be the same  as the ordinary RN metric. 
Nonetheless, the problem with such stance is that it completely ignores the context which brought about the metric  \eqref{ncrn} and in which it was derived.
Regarding the context in this concrete example, imagine that 
 we have two spacetimes, the background $(M,g_{ab})$ with RN metric $g_{ab}$, where we place NC scalar field, and ``effective'' spacetime $(M,g'_{ab})$ with the effective metric $g'_{ab}$. Unfortunately, as the whole setting (background spacetime and NC action) is prepared in a specific coordinate system, we cannot easily transform components of the effective metric $g'_{ab}$ without going back to the origin of this construction. At best, coordinate changes as the one above, may be trusted at infinity.

Additional point in this case  is that the coordinates are noncommutative  and the partial derivatives are also noncommutative. In particular, 
\begin{equation}
    \partial_{\tilde{r}}=\partial_{r}-\frac{A}{r^2}\partial_{\varphi}.
\end{equation}
From these reasons, it is clear that the new geometry will have nontrivial NC effects up to the first order in NC deformation parameter $a,$ contrary to the argumentation made around 
(\ref{coordtr}) and after. Not a bit less important is that the coupling of the NCRN metric with other fields makes a huge difference in comparison with a situation when this metric is taken alone  and studied as an isolated entity.  This is where the importance  of the duality symmetry  and a validity of the corresponding requirement  \eqref{dd}
comes into play. Namely, after the coordinate transformation leading to (\ref{coordtr}) the duality does not hold anymore. 

The latter argument is readily confirmed in references \cite{Dimitrijevic-Ciric:2022vvl},\cite{rnqnm},\cite{DimitrijevicCiric:2019hqq}, where the metric \eqref{ncrn}
was   coupled to the spin $1/2$ field and scalar field, respectively. We point out that already at the linear order in the deformation parameter these couplings result  with the QNM spectra that  differ from the corresponding QNM spectra when the same fields are coupled to the ordinary RN metric. In this way, the assertion that at the first order in deformation the NCRN metric is essentially  the same as the RN metric directly contradicts with findings in \cite{rnqnm},\cite{DimitrijevicCiric:2019hqq} where it was explicitly shown that scalar perturbations of RN and NCRN give rise to different 	QNM spectra already at linear order in the deformation parameter. As well, this assertion is in contradiction with the findings in \cite{Dimitrijevic-Ciric:2022vvl}
which show that the  spin $1/2$ field perturbations of RN and NCRN are governed with different equations of motion.



In summary,
when talking about physical properties of the NCRN metric \eqref{ncrn},
we may conclude by saying that observed only for itself, outside of the context in which it was derived, it appears to be just the RN metric in
different coordinates.  However, what brings something new to this metric and its consequences for physics is
when it couples to other types of fields, for example the scalar, spinor  and  gauge field.


\bigskip

\noindent{\bf Acknowledgment}\\ 
We would like to thank Kumar Gupta for fruitful discussion and
useful comments. This research was supported by the Croatian Science Foundation Project  IP-2020-02-9614
{\it{Search for Quantum spacetime in Black Hole QNM spectrum and Gamma Ray Bursts}}. The work of M.D.C.
and N.K. is supported by project 451-03-9/2021-14/200162 of the Serbian Ministry of Education and
Science.  This work
is partially supported by ICTP-SEENET-MTP Project NT-03 ”Cosmology-Classical
and Quantum Challenges” in frame of the Southeastern European Network in Theoretical and Mathematical Physics and the COST action CA23130 {\it{Bridging high and low energies in search of quantum gravity (BridgeQG).}}


\appendix
\section{Second order corrections to the effective metric}

In this appendix we describe the challenges one encounters when trying to deduce the form of the effective metric
in higher orders of deformation. In order to make a construction in higher orders possible, one has to extend the existing framework and redefine the coefficients of affine connection $\Gamma^\rho_{\mu\nu}$  in such a way that besides the ordinary Christoffels $\{^\rho_{\mu\nu}\}$ they also  include the contorsion $K^\rho_{\mu\nu}$ and the nonmetricity $C^\rho_{\mu\nu}$.
$$\Gamma^\rho_{\mu\nu}=\{^\rho_{\mu\nu}\}+\frac{1}{2}C^\rho_{\mu\nu}+K^\rho_{\mu\nu}.$$
Nonmetricity and contorsion are the symmetric and antisymmetric parts of the connection, respectfully.

It is readily seen that the  second order effective metric obtained by adding the contribution (\ref{gphiphi2a})
to the first order metric (\ref{1stordereffmetric}) can easily account for the terms with second derivatives   in the equation (\ref{2ndorderequation}). However, the issue with first order derivatives becomes more involved.
It appears that the only way to account for these redundant first derivative terms is to extend the connection as described above so that nonmetricity and contorsion may absorb these  terms. 
More precisely, from  (\ref{dd})  it can be seen that the first  derivative terms have a form
\begin{equation}\label{razvoj}
-g^{\mu\nu}\Gamma^\rho_{\mu\nu}\partial_\rho \Phi=-g^{\mu\nu(0)}\Gamma^{\rho(2)}_{\mu\nu}\partial_\rho\Phi-g^{\mu\nu{1}}\Gamma^{\rho(1)}_{\mu\nu}\partial_\rho\Phi-g^{\mu\nu(2)}\Gamma^{\rho(0)}_{\mu\nu}\partial_\rho\Phi
\end{equation}
Here all necessary nonzero Christoffels may be calculated from (\ref{gphiphi2a}) and (\ref{gphiphi2b}) to give
\begin{eqnarray}
    &&\{^r_{r r}\}^{(2)}=-(aqQ)^2\frac{f\sin^2\theta}{4r^3},\quad \{^r_{\varphi \varphi }\}^{(2)}=(aqQ)^2\frac{ff'\sin^4\theta}{8},\nonumber\\
    && \{^{\theta}_{r r}\}^{(2)}  =-(aqQ)^2\frac{\sin\theta\cos\theta}{4r^4}, \quad   \{^{\theta}_{\varphi \varphi }\}^{(2)}=-(aqQ)^2\frac{f\cos\theta\sin^3\theta}{2r^2},\nonumber\\
    &&\{^{\theta}_{r \varphi }\}^{(1)}=-(aqQ)\frac{\sin\theta\cos\theta}{2r^2},\quad    \{^{\theta}_{\varphi \varphi }\}^{(0)}=-\sin\theta\cos\theta,\nonumber\\
    &&\{^r_{\varphi \varphi }\}^{(0)}=-rf\sin^2\theta.\nonumber
\end{eqnarray}

Interestingly, as the contorsion needs to be  antisymmetric in the  last two indices, all its contributions to (\ref{razvoj}) will vanish as they need to be contracted with the inverse metric tensor $g^{\mu\nu}$ which is symmetric. This means that the only terms that may annihilate
the first derivative corrections in (\ref{razvoj}) 
\begin{equation}
    (aqQ)^2\frac{f\sin\theta\cos\theta}{4r^4}\partial_\theta\Phi+(aqQ)^2(f'-\frac{6f}{r})\frac{f\sin^2\theta}{8r^2}\partial_r\Phi,
\end{equation}
 are those that involve  components of the nonmetricity tensor.

 There are several ways how one can remove unwanted first derivative terms:
\begin{itemize}
    \item Nonmetricity may be introduced as a  first order deformation, so that we may demand
    $$-g^{r\varphi(1)}C^{\theta(1)}_{r\varphi}=(aqQ)^2\frac{f\sin\theta\cos\theta}{4r^4}\Rightarrow C^{\theta(1)}_{r\varphi}=(aqQ) \frac{\sin\theta\cos\theta}{4r^2}$$

    $$-g^{r\varphi(1)}C^{r(1)}_{r\varphi}=(aqQ)^2(f'-\frac{6f}{r})\frac{f\sin^2\theta}{8r^2}\Rightarrow C^{r(1)}_{r\varphi}=(aqQ)(f'-\frac{6f}{r})\frac{\sin^2\theta}{8}.$$
Thereof the components of nonmetricity immediately follow.
\item Nonmetricity may be introduced as a  second order deformation, so that we may demand
 $$-\frac{1}{2}g^{rr(0)}C^{\theta(2)}_{rr}=(aqQ)^2\frac{f\sin\theta\cos\theta}{4r^4}\Rightarrow C^{\theta(2)}_{rr}=-2(aqQ)^2\frac{\sin\theta\cos\theta}{4r^4},$$

    $$-\frac{1}{2}g^{rr(0)}C^{r(2)}_{rr}=(aqQ)^2(f'-\frac{6f}{r})\frac{f\sin^2\theta}{8r^2}\Rightarrow C^{r(2)}_{rr}=-2(aqQ)^2(f'-\frac{6f}{r})\frac{\sin^2\theta}{8r^2}.$$

    The other possibility is

     $$-\frac{1}{2}g^{\varphi\varphi(0)}C^{\theta(2)}_{\varphi\varphi}=(aqQ)^2\frac{f\sin\theta\cos\theta}{4r^4}\Rightarrow C^{\theta(2)}_{\varphi\varphi}=-(aqQ)^2\frac{f\sin^3\theta\cos\theta}{2r^2},$$

    $$-\frac{1}{2}g^{\varphi\varphi(0)}C^{r(2)}_{\varphi\varphi}=(aqQ)^2(f'-\frac{6f}{r})\frac{f\sin^2\theta}{8r^2}\Rightarrow C^{r(2)}_{\varphi\varphi}=-(aqQ)^2(f'-\frac{6f}{r})\frac{f\sin^4\theta}{4}.$$

   We may conclude that the way of implementing nonmetricity, which the framework we work in allows,  is certainly not unique. It would be interesting to understand more deeply the physical consequences of the effective non-metric geometry. Also, we point out that while the equation of motion for the NC scalar field is exact,  all other results such as the components of the metric and nonmetricity tensors  are not exact. They are instead perturbative and given only up to a second order in the NC parameter $a$ (i.e. they have higher order corrections).
 \end{itemize}


\end{document}